\documentclass[12pt]{article}

\usepackage{amsthm,graphicx,cite,lscape,float}
\usepackage{epsf,latexsym}
\usepackage{amssymb,amsmath}

\usepackage{epsf,latexsym}
\usepackage{amssymb,amsmath,caption}
\newcommand{\fpm}{f_{\mbox{\scriptsize $\parallel$}}}

\begin{document}

\title{\bf Reply to Comment on ``Comparison of the Lateral Retention Forces on
Sessile, Pendant, and Inverted Sessile Drops''}

\author{Rafael de la 
Madrid,\footnote{E-mail: \texttt{rafael.delamadrid@lamar.edu}} \ 
Fabian Garza,\footnote{Current address: Department of Physics, 
Texas A\&M University, College Station, TX 77843} \
Justin Kirk, 
Huy Luong,\footnote{Current address: Sage Automation, Beaumont, TX 77705} \\
Levi Snowden, 
Jonathan Taylor,$^{\dagger}$ 
Benjamin Vizena\footnote{Current address: METECS, Houston, TX 77289} \\ [2ex]
\small{\it Department of Physics, Lamar University,
Beaumont, TX 77710} }

\date{\small{\today}}


\maketitle

\begin{abstract}

\noindent We address the issues raised in [R.~Tadmor et al., 
{\it Langmuir}~{\bf 2020},
{\it 36}, 475-476]. In 
particular, we explain why we did not use Tadmor's theory
to explain our results.
\end{abstract}

\section{Introduction}

We have recently reported an experiment~\cite{LANGMUIR19} that 
provides a possible explanation of the results of Ref.~\cite{TADMOR09}. In 
a Comment~\cite{COMMENT}, 
Tadmor {\it et al.}~compare the results of Refs.~\cite{TADMOR09}
and~\cite{PP} 
with ours, and they raise some issues. In this Reply, 
we would like to clarify some of those issues.

Before we address those issues, it is important to note that
in Refs.~\cite{LANGMUIR19} and~\cite{TADMOR09}, the drops  
were placed on a horizontal surface, and an increasing centrifugal 
force parallel to the surface made them slide. In 
Ref.~\cite{PP}, the drops were placed on a vertical surface, the 
centrifugal force was perpendicular to the surface, and the (constant) 
weight was the force that made the drops slide. Hence, we do not think that
a comparison of our results with those of Ref.~\cite{PP} is very 
meaningful~\cite{NOTE1}, and in this Reply we will mainly focus on the 
comparison of our results with the original paper~\cite{TADMOR09}.

\section{Experimental Aspects}

As pointed out in Comment~\cite{COMMENT}, to determine the ``onset of 
motion'' of the drops, we established the criterion that the receding edge of 
the drop had to be moving at around $0.09$~mm/s. There are two important 
reasons why we chose such criterion. 
First, in our water-PMMA system, the drops exhibit a short period of stick-slip
motion as they transition from rest to motion, and therefore there is a certain
degree of ambiguity to determine when the onset of the motion occurs. This
ambiguity is not unique to our system, as can be seen for example
in Figure~6 of Ref.~\cite{SR}. We 
realized that whenever the receding edge was moving at around 0.09~mm/s, the
drop would be past the stick-slip period, and it seemed 
to us like a good criterion to define onset of motion.

There is a second, and more important, reason why we chose 
0.09~mm/s. In Ref.~\cite{TADMOR09}, the motion of the drops was determined 
by taping a ruler to the computer monitor and 
writing on the screen with a pencil to mark the position of the drop for
each frame in the video taken by the camera of the
Centrifugal Adhesion Balance (CAB)~\cite{TAYLOR}. Thus, essentially, in 
Ref.~\cite{TADMOR09} the onset of the motion was determined by visual
inspection of the drop as it appears on a computer 
screen. However, the stick-slip motion of the drop mentioned above is usually
undetectable to the naked eye. We realized that for our naked eyes to
notice motion by visual inspection of the computer screen, the drop
would have to be moving at about 0.09~mm/s.  Hence, for practical purposes,
our criterion to determine the onset of motion of a drop is the same as in 
the original paper~\cite{TADMOR09}.

In addition, our criterion is more consistent
than that of Ref.~\cite{TADMOR09}. In 
the CAB used in the original paper~\cite{TADMOR09}, the starting times of 
the video and the motor were not synchronized~\cite{KEN}, and in fact the 
raw data files for the centrifugal force and the video have different starting 
times~\cite{TAYLOR}. Those times were synchronized afterwards, and it 
is not clear how well such synchronization can be done~\cite{KEN}. 
In our experiment, those times were synchronized by turning on
the light and the motor at the same time~\cite{GRIEGOS1,GRIEGOS2,GRIEGOS3}.

Analyzing the motion of drops by following them on a computer
screen is very time consuming, and the data of Ref.~\cite{TADMOR09}
are based on a few runs. In fact, the plots have no error bars, and the results
are reported without any uncertainties. Our results are based on 624 runs,
and we believe that our data have statistical significance and that only an 
unknown systematic error could upset our conclusions.

Contrary to the CAB, our experimental setup provides a top view 
of the drops that can be used to measure their width. As we will see below, 
the width was critical in
determining which theory would best fit the experimental results.

\section{Theoretical Aspects}

\begin{figure}[h!]
\begin{flushleft}
              \epsfxsize=8cm
              \epsffile{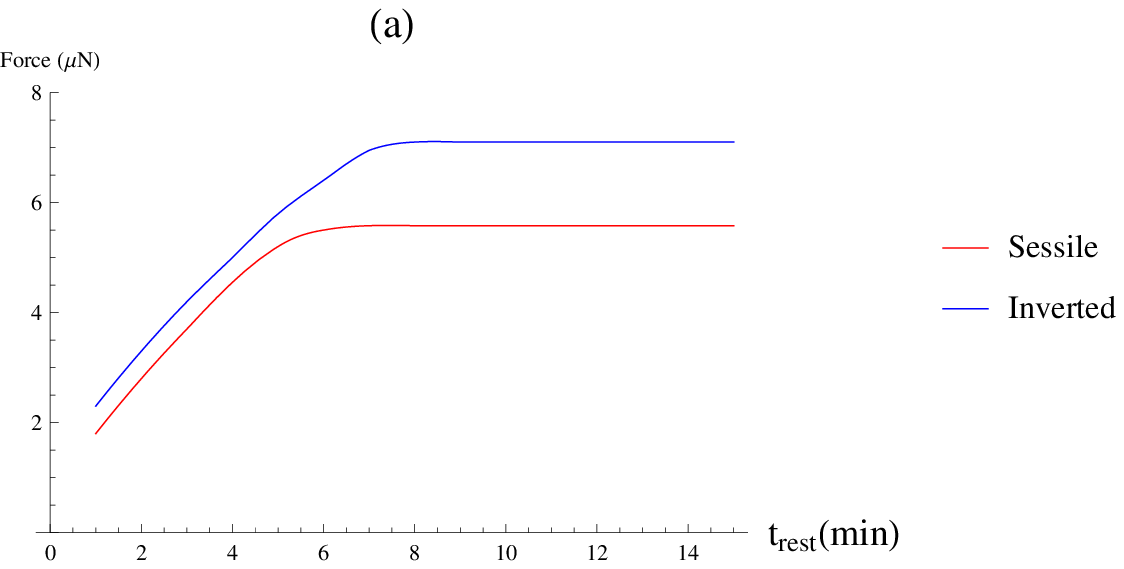} \hskip1cm
           \epsfxsize=6cm \epsffile{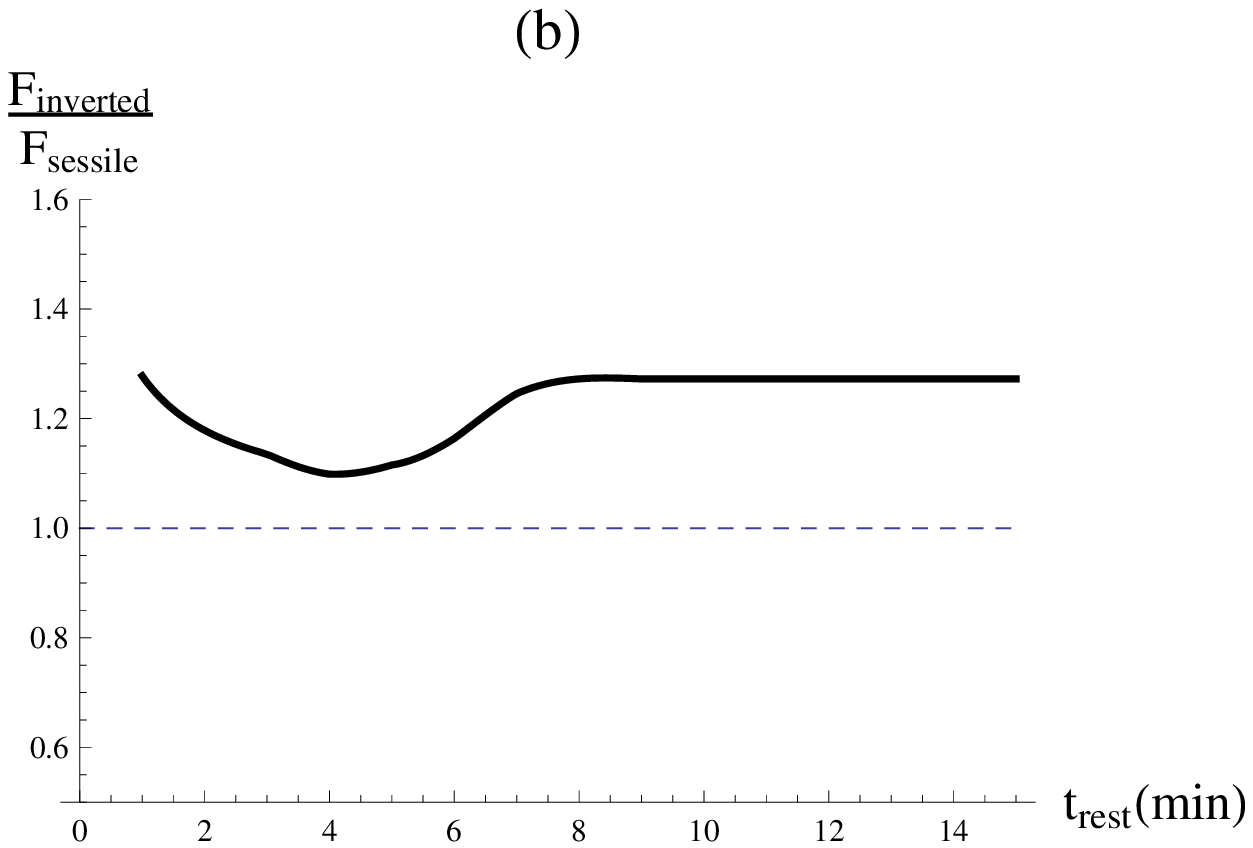}
\end{flushleft}
\caption{(a) Retention force on sessile and inverted drops versus the 
resting time, and (b) ratio $\frac{F_{\rm inverted}}{F_{\rm sessile}}$. Data 
extracted from Ref.~\cite{TADMOR09}.}
\label{fig:spto}
          \end{figure}

The main results of the original paper~\cite{TADMOR09}, which are
displayed in Fig.~\ref{fig:spto},  
are that ({\it i}) the retention force on an inverted drop~\cite{NOTE3} is 
greater than on a sessile drop for any resting time, 
$F_{\rm inverted}>F_{\rm sessile}$, 
and ({\it ii}) the retention force (for both sessile and inverted drops) 
increases with the resting time until it reaches a plateau. The data of 
Fig.~\ref{fig:spto} have been extracted from Figure~3 of the original 
paper~\cite{TADMOR09}.

To explain those results, Tadmor proposed the following retention 
force~\cite{PP,SR,MIS},
\begin{equation}
   \fpm = \frac{4 \gamma ^2 \sin \theta_{\rm still}}{G_{\rm s}} 
        (\cos \theta_{\rm r}  - \cos \theta _{\rm a}) \, , 
     \label{tadthe}
\end{equation}
where $\gamma$ is the liquid-vapor surface tension, $\theta _{\rm still}$ is 
the contact angle that the drop adopted during the 
``still period''~\cite{TADMOR09},
$G_{\rm s}$ is the interfacial modulus, and $\theta _{\rm a}$ ($\theta _{\rm r}$)
is the advancing (receding) angle. In our work~\cite{LANGMUIR19}, we used the 
standard expression
\begin{equation}
      \fpm = k \gamma w \left( \cos \theta _{\rm r}- \cos \theta _{\rm a} \right)
        \, ,
   \label{fpw}
\end{equation}
where $w$ is the width of the drop and $k$ is a shape factor.

A critical difference between Tadmor's theory and the
standard approach is that in Eq.~(\ref{tadthe}) the lateral
retention force does not depend on the size\cite{MIS} (or shape) 
of the drop, whereas in 
Eq.~(\ref{fpw}) there is a size (and shape) dependence. Because our 
experimental retention force was size dependent, we could simply not 
use Eq.~(\ref{tadthe}) to fit our experimental data, and hence resorted
to Eq.~(\ref{fpw}). Furthermore, our data showed that the lateral retention
force is (within the accuracy of our experiment) proportional to the width 
of the drop~\cite{NOTE4}, in agreement with the prediction of Eq.~(\ref{fpw}).

A key ingredient of our experiment was that what we call an 
``inverted drop''~\cite{NOTE3} is different from a ``pendant drop.'' According 
to Eq.~(\ref{tadthe}), the retention force on a pendant drop
should be greater than on an inverted drop, 
$F_{\rm pendant} >F_{\rm inverted}$, but our experimental results showed 
the opposite, $F_{\rm pendant} <F_{\rm inverted}$, in agreement with the
predictions of Eq.~(\ref{fpw}) -- yet another reason why we used
Eq.~(\ref{fpw}), rather than Eq.~(\ref{tadthe}), to fit 
our experimental data~\cite{NOTE8}.

Although not directly related to the issues raised in 
Comment~\cite{COMMENT}, we would like to mention two interesting aspects 
of Fig.~\ref{fig:spto}. First, even though the experimental retention forces
of Fig.~\ref{fig:spto} depend on the resting time, the retention force of
Eq.~(\ref{tadthe}) is resting-time {\it independent}: $G_{\rm s}$ has been 
reported to be resting-time independent~\cite{SR}, the advancing and receding 
angles are supposedly resting-time independent~\cite{TADMOR09}, 
and $\theta _{\rm still}$ should be 
resting-time independent if evaporation is suppressed. Thus, one cannot
use Eq.~(\ref{tadthe}) to fit the actual data of the original paper for all
resting times. Second, we can see in Fig.~\ref{fig:spto}(b) that the ratio 
$\frac{F_{\rm inverted}}{F_{\rm sessile}}$ varies in a complicated way: It 
starts with a large value (around 1.27) at $t_{\rm resting}\simeq 1$~min, 
drops to a minimum of
about 1.1 at $t_{\rm resting}\simeq 4$~min, and finally approaches the
plateau value of 1.27 after $t_{\rm resting}\simeq 8$~min. Equation~(\ref{tadthe})
cannot explain such complicated
behavior.

\section{Specific Issues}

We are now in a position to address the three specific issues raised in 
Comment~\cite{COMMENT}.

{\bf 1.~Considering the Drops in Motion vs.~Considering Their Onset of
Motion}

In their Comment~\cite{COMMENT}, Tadmor~{\it et al.}~state that
in the original paper~\cite{TADMOR09}, the onset of the motion was
determined by monitoring the advancing edge, instead of the receding 
edge. Several
comments are in order. First, the drop as a whole does not move until
the receding edge moves. Usually, the motion of the advancing edge
is called spreading~\cite{GRIEGOS1,GRIEGOS2,GRIEGOS3}. Second, from the
frames of drops in Figure~2b of the original paper~\cite{TADMOR09}, one
has the impression that the onset of the motion was determined by monitoring
the receding edge. Third, in a recent publication~\cite{SR}, they have
explicitly defined the onset of motion in terms of the receding edge. Fourth, 
the contact angles at the instant
the front edge starts moving are not the same as at the instant the rear edge 
starts moving~\cite{GRIEGOS1,GRIEGOS2,GRIEGOS3}. Because 
$(\cos \theta_{\rm r}  - \cos \theta _{\rm a})$ is very sensitive to small
changes in $\theta _{\rm a}$ and $\theta _{\rm r}$, the theoretical value of 
the retention force at the moment the front edge starts moving can be very 
different from its value at the moment the rear edge starts 
moving. Fifth, our drops do not move at constant speed 
0.09~m/s. As the centrifugal force continues growing, the drops continue 
speeding up. As explained above, we used that speed to determine an onset
of motion that would be equivalent to visually monitoring the motion of 
the drops on a computer screen~\cite{TADMOR09}.

As Tadmor {\it et al.}~point out, the shape factor $k$ is for practical
purposes a free
parameter of our fit that can only be obtained once we measure the 
other quantities in Eq.~(\ref{fpw})~\cite{NOTE6}. However, the
interfacial modulus $G_{\rm s}$ of Eq.~(\ref{tadthe}) is also a quantity that
can only be obtained once the other quantities of Eq.~(\ref{tadthe})
have been measured. In fact, if we used
Eq.~(\ref{tadthe}) to fit our data, $G_{\rm s}$ would be a free parameter
(fudge factor) of the fit.

{\bf 2.~Resting Time Effect.}

We agree with Tadmor {\it et al.}~that our results do not take into account 
the effect of resting time, but this does not affect our conclusions. The 
reasons are the following. First, as can be seen in Fig.~\ref{fig:spto}(b),
the ratio $\frac{F_{\rm inverted}}{F_{\rm sessile}}$ is always greater than unity, 
and therefore one does not have to wait until the plateau region to observe 
that $\frac{F_{\rm inverted}}{F_{\rm sessile}}>1$. Second, in Fig.~\ref{fig:spto}(b)
the ratio $\frac{F_{\rm inverted}}{F_{\rm sessile}}$ is about the same 
for short (around 1 minute) and for long (around 8 minutes) resting
times, so one does not necessarily have to wait a long time to observe 
a large value of $\frac{F_{\rm inverted}}{F_{\rm sessile}}$. Third, 
Eqs.~(\ref{tadthe}) and (\ref{fpw}) are resting-time independent, and 
hence any experimental work on the resting time effect 
cannot be fitted with either Eq.~(\ref{tadthe}) or (\ref{fpw}).

{\bf 3.~Drop's Size and Bond Number.}

As discussed above, Eq.~(\ref{tadthe}) applies to all volumes (it is 
volume independent), and hence it should be true also for the volume
range we used (15-100~$\mu$L). Hence, if Eq.~(\ref{tadthe}) was the
general formula for the retention force~\cite{MIS}, it would produce a 
good fit for 
all volumes~\cite{NOTE7}, independently of whether our main result 
($F_{\rm inverted}>F_{\rm sessile}>F_{\rm pendant}$) is also true for small volumes.

In their Comment~\cite{COMMENT}, the authors mention some quantities
(Laplace pressure, torque, center of mass, contact area) that may
influence the retention force. However, since none of those quantities appear
in Eqs.~(\ref{tadthe}) or~(\ref{fpw}), their influence on the retention
force cannot be checked against experimental data, and therefore we will
not discuss their bearing on our results.

\section{Conclusion}

Using a criterion to determine the onset of the motion of the
receding edge of a drop that would
yield the same results as the one used in the original 
paper~\cite{TADMOR09}, we determined experimentally that 
$F_{\rm inverted}>F_{\rm sessile}>F_{\rm pendant}$. Because Tadmor's theory, 
Eq.~(\ref{tadthe}), is inadequate to fit our data, we resorted to the 
standard, simple theory, Eq.~(\ref{fpw}), which provides a reasonably good fit. 

Our results~\cite{LANGMUIR19} did not account for the influence that
quantities such as resting time, Laplace pressure, contact area or torque 
may have on the retention force. However, because (to the best of our knowledge)
there is no theory that takes into account such influence and that can be 
used to fit experimental data, we refrain from speculating on how
those quantities may affect our results.


\begin{thebibliography}{99}
\bibitem{NOTE1} In addition, an alternative explanation of the results of 
Ref.~\cite{PP}, with a proper discussion of the systematics of the
experiment, would need a separate paper.
\bibitem{NOTE3} In the original paper\cite{TADMOR09} and
in Comment,\cite{COMMENT} inverted drops are
called pendant drops.
\bibitem{NOTE4} This result was possible thanks to the
top view of the drop, which the CAB lacks.
\bibitem{NOTE6} To avoid using the shape factor $k$, we would have to 
measure the contact angle along every point of the triple line. A
successful such measurement can be found in Ref.~\cite{FEAT}. We 
thank Stefan Zitz for pointing out Ref.~\cite{FEAT} to us.
\end{thebibliography}
\end{document}